\documentclass[10pt,a4paper,conference,onecolumn]{IEEEtran}
\newcommand{\KBERcode}[1]{\textsl{#1}}
\usepackage{graphicx}
\usepackage{algorithm}
\usepackage{algpseudocode}
\usepackage{psfrag}
\usepackage{alltt}
\usepackage{amssymb}
\usepackage[english]{babel}
\usepackage[noadjust]{cite}
\begin{document}
\bibliographystyle{IEEEtran}

\title{FreeBSD Mandatory Access Control Usage for Implementing Enterprise Security Policies}

\author{\IEEEauthorblockN{Kirill~Bolshakov\hspace{5cm}Elena~Reshetova}
\IEEEauthorblockA{\hspace{4mm}kirill.bolshakov@ieee.org\hspace{4cm}elena.reshetova@gmail.com}
Saint-Petersburg State University of Aerospace Instrumentation
}

\maketitle

\begin{abstract}
FreeBSD was one of the first widely deployed free operating systems
to provide mandatory access control. It supports a number of classic
MAC models. This tutorial paper addresses exploiting this implementation
to enforce typical enterprise security policies of varying complexities.
\end{abstract}

\section{Introduction}
Security needs of organizations are becoming more and more 
sophisticated nowadays. Most general-purpose operating 
systems (GPOS) provide access control policies to meet 
these needs. There are cases when the traditionally deployed 
Discretionary Access Control (DAC) rules are not sufficient: 
they tend to quickly become unmanageable in the case of 
large installations, and also are not enough for
controlling information flows. This is when the Mandatory 
Access Control (MAC) comes in: it provides for better 
manageability and directly targets the information flows. 
In their turn, the information flows address the confidentiality and 
integrity needs of information security within an organization. 
Until very recently, the GPOSes tended to provide various 
flavors of DAC only. The FreeBSD OS \cite{KBER:FreeBSD} was one of the first 
widely deployed open source GPOSes to support MAC \cite{DBLP:conf/usenix/Watson01,DBLP:conf/usenix/WatsonMVF03}. 
In this paper, 
a number of organizational policy examples are implemented in
the environment of the FreeBSD MAC.

The authors
strongly believe that in order to implement a sound MAC policy
it is important to understand MAC's mathematical foundations.
These foundations were set by Denning in \cite{DBLP:journals/cacm/Denning76}.
There also exists a terminology confusion between MAC and LBAC 
(lattice-based access control). These models are the same, because
MAC security labels \cite{KBER:Orange85} directly correspond to
security classes of lattice-based models (this was also pointed to
by Sandhu \cite{DBLP:journals/computer/Sandhu93} and Osborn \cite{DBLP:conf/sacmat/Osborn02}).

Let us first address the definition of the information flow. 
According to Denning and Sandhu,
the security policies regulate how the information ``flows from 
one object to another''. A typical object is 
a shared memory segment, a file system object or a network 
packet. Obviously, controlling the information flows is important
to prevent the leakage of the confidential information, the one
usually sought by insiders. Another goal is the forgery 
prevention, so that
no untrusted reports are ever submitted to the top level
of the organization hierarchy, and no top-ranking company
officers take any unchecked or untrusted information into
account during decision making.

To implement the information flow control, every object is assigned a 
security label (also called 
a security classification), implemented by the FreeBSD file label. 
When the information flows from one object 
into another, \emph{an information flow} from the security class of the 
first object to the security class of the second one also takes 
place. Whether the information flow is allowed is regulated by the 
relation between the object security classes.
The subjects are the entities performing the information transfer
between the objects. In our case, a subject appears when a user
logs in to the system and is assigned a set of privileges. As we are 
considering MAC, the set of privileges is rendered as security clearance.
It is implemented by the FreeBSD user label.

This paper is organized as follows. In the next section an example
of an organization and its document flow is described. The following
sections implement organization's information security goals, which gradually 
increase in complexity. The information security goals
specify the target effect: preserving data and process integrity, restricting access
to the confidential information, or implementing a consulting services policy.
For every security goal, a corresponding classic MAC model or a
combination of them is chosen. The models are then implemented in the
FreeBSD MAC framework.

\section{The Example}

\begin{figure}[ht]
\centering
\includegraphics{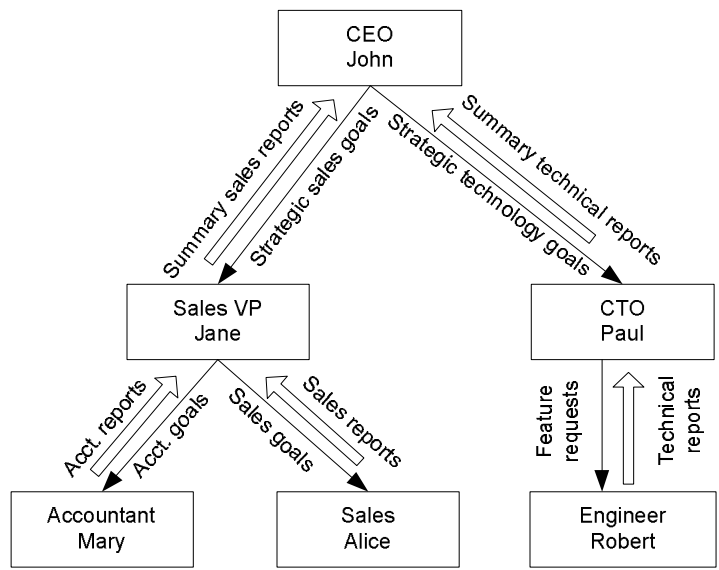}
\caption{Example Organization Chart and Document Flows}
\label{KBER_InfFlows}
\end{figure}

The example organization we consider is a very small technology company. There are
only six positions within the organization. The organization chart and the exchange 
of files corresponding to the information flows are shown in Fig.~\ref{KBER_InfFlows}. 
The document flow is bidirectional. 

The file system folder layout reflects this, and the folders are listed in 
Table \ref{tab:FolderLayout}. Operations will be applied to the files in these folders. The \KBERcode{Temp} folder is
used for file exchange purposes. The administrator may consider configuring this
folder with the ``sticky'' attribute (so that the files could be deleted by their
owners only). The \KBERcode{``U''} (from ``Untrusted'') folders are used by Mary, Alice and Robert for delivering
documents to Jane and Paul, respectively. After the managers review these untrusted
documents and deem them trustworthy, the files may be copied to the corresponding no-prefix folders
while increasing their security class. This makes these documents available 
for further processing to the higher levels of hierarchy: for example, they may be included into
Jane's or Paul's reports for John.

In FreeBSD, the MAC labels are assigned to the user accounts via the notion of \emph{login classes.}
The mapping of user accounts to their login classes is given in the \KBERcode{master.passwd} file, 
where the user accounts are traditionally stored. The login classes 
are listed alongside the group information. The login classes themselves are 
specified in the \KBERcode{login.conf} file. In the examples, we will list only the 
MAC-related configuration of the login classes. The MAC labels used in FreeBSD have the
syntax of \KBERcode{policy1/qualifier1,policy2/qualifier2,...}, where the policy component
describes the policy module (e.g., Biba or MLS), and the qualifier describes a corresponding 
grade, compartments, and a combined grade-compartment range \cite{KBER:Handbook}.

\begin{table}[ht]
\centering
\caption{Folder Layout}\label{tab:FolderLayout}
\begin{tabular}{|l|l|}\hline
AccountingGoals          & StrategicTechnologyGoals \\
AccountingReports                  & SummarySalesReports \\
SummaryTechnicalReports  & TechnicalReports\\
FeatureRequests                            & Temp\\
SalesGoals                                                       & UAccountingReports\\
SalesReports                                             & USalesReports\\
StrategicSalesGoals              & UTechnicalReports\\
\hline
\end{tabular}
\end{table}

In the following sections the increasingly complex information flow 
control goals are considered. The goals are then translated into 
a MAC policy sufficient to fulfill them, and then into a
FreeBSD implementation.

\section{Protecting the Integrity}

Suppose that the organization needs to protect itself from making
decisions based on the untrusted data. That is, John should not
make decisions based on the data directly received from Alice or
Robert.  This corresponds to the John's subject (which is his user login) 
having a higher integrity label: he
should be prevented from inadvertently reading a document of the
lower integrity label than his own. Also, John's requests should 
propagate through the organization unchanged: his documents
cannot be tampered with by the subjects with a lower integrity.
In the information flow sense, the information is allowed to flow
from the higher integrity to the lower integrity objects, but the
reverse flow is disallowed. This is the Biba model with the liberal
$\star$-property \cite{DBLP:journals/computer/Sandhu93}.
It is directly supported in FreeBSD by the \KBERcode{mac\_biba} module.
The assignment of the labels to the files in the case of the Biba
model support will look like shown in Table \ref{tab:FileLabelsForBibaModel},
and the user label assignment is shown in Table \ref{tab:UserLabelsForBibaModel}.
The \KBERcode{Temp} folder is excluded from the policy by setting its grade to
\KBERcode{equal,} because it serves
as a document exchange location.

\begin{table}[ht]
\parbox[t]{0.29\textwidth}{\caption{Files: Biba}\label{tab:FileLabelsForBibaModel}}
\hfill
\parbox[t]{0.21\textwidth}{\caption{Files: Biba and MLS}\label{tab:FileLabelsForMLSModel}}
\hfill
\parbox[t]{0.35\textwidth}{\caption{Files: Biba, MLS and Compartments}\label{tab:FileLabelsForCompartments}}
\begin{tabular}{|l|l|}\hline
AccountingGoals          & biba/2\\
AccountingReports        & biba/5\\
SummaryTechnicalReports  & biba/10\\
FeatureRequests          & biba/2\\
SalesGoals               & biba/2\\
SalesReports             & biba/5\\
StrategicSalesGoals      & biba/5\\
StrategicTechnologyGoals & biba/5\\
SummarySalesReports      & biba/10\\
TechnicalReports         & biba/5\\
Temp                     & biba/equal\\
UAccountingReports & biba/2\\
USalesReports & biba/2\\
UTechnicalReports & biba/2\\
\hline
\end{tabular}
\hfill
\begin{tabular}{|l|l|}\hline
AccountingGoals & biba/2,mls/low\\
AccountingReports & biba/5,mls/low\\
SummaryTechnicalReports & biba/10,mls/50\\
FeatureRequests & biba/2,mls/50\\
SalesGoals & biba/2,mls/low\\
SalesReports & biba/5,mls/low\\
StrategicSalesGoals & biba/5,mls/low\\
StrategicTechnologyGoals & biba/5,mls/50\\
SummarySalesReports & biba/10,mls/low\\
TechnicalReports & biba/5,mls/50\\
Temp & biba/equal,mls/equal\\
UAccountingReports & biba/2,mls/low\\
USalesReports & biba/2,mls/low\\
UTechnicalReports & biba/2,mls/50\\
\hline
\end{tabular}
\hfill
\begin{tabular}{|l|l|}\hline
AccountingPlans & biba/2,mls/50:1\\
AccountingReports & biba/5,mls/50:1\\
CommonTechnicalReports & biba/10,mls/50:2\\
FeatureRequests & biba/2,mls/50:2\\
FinancialReports & biba/10,mls/50:1\\
SalesPlans & biba/2,mls/50:1\\
SalesReports & biba/5,mls/50:1\\
StrategicMarketingGoals & biba/5,mls/50:1\\
StrategicTechnologyGoals & biba/5,mls/50:2\\
TechnicalReports & biba/5,mls/50:2\\
Temp & biba/equal,mls/equal\\
UAccountingReports & biba/2,mls/50:1\\
USalesReports & biba/2,mls/50:1\\
UTechnicalReports & biba/2,mls/50:2\\
\hline
\end{tabular}
\vspace{7mm}\\
\parbox[t]{0.19\textwidth}{\caption{Users: Biba}\label{tab:UserLabelsForBibaModel}}
\hfill
\parbox[t]{0.30\textwidth}{\caption{Users: Biba and MLS}\label{tab:UserLabelsForMLSModel}}
\hfill
\parbox[t]{0.35\textwidth}{\caption{Users: Biba, MLS and Compartments}\label{tab:UserLabelsForCompartments}}\\
\begin{tabular}{|l|l|}\hline
John & biba/10(10-10)\\
Jane & biba/5(2-10)\\
Paul & biba/5(2-10)\\
Alice & biba/2(2-2)\\
Mary & biba/2(2-2)\\
Robert & biba/2(2-2)\\
\hline
\end{tabular}
\hfill
\begin{tabular}{|l|l|}\hline
John & biba/10(10-10),mls/100(100-100)\\
John.Sales & biba/10(10-10),mls/low(low-low)\\
John.Engineering & biba/10(10-10),mls/50(50-50)\\
Jane & biba/5(2-10),mls/low(low-low)\\
Paul & biba/5(2-10),mls/50(50-50)\\
Alice & biba/2(2-2),mls/low(low-low)\\
Mary & biba/2(2-2),mls/low(low-low)\\
Robert &  biba/2(2-2),mls/50(50-50)\\
\hline
\end{tabular}
\hfill
\begin{tabular}{|l|l|}\hline
John & biba/10(10-10),mls/100:1+2(100-100:1+2)\\
John.Lower & biba/10(10-10),mls/50:1+2(50-50:1+2)\\
Jane & biba/5(2-10),mls/50:1(50:1-50:1)\\
Paul & biba/5(2-10),mls/50:2(50:2-50:2)\\
Alice & biba/2(2-2),mls/50:1(50:1-50:1)\\
Mary & biba/2(2-2),mls/50:1(50:1-50:1)\\
Robert & biba/2(2-2),mls/50:2(50:2-50:2)\\
\hline
\end{tabular}
\end{table}

While the Biba policy enforces the integrity, it also disallows the reverse
flow of information. It is obvious that the management should be aware
of what is going on in the managed departments, so the reverse flow is required
for the considered example.
This is achievable in FreeBSD by adding \emph{ranged labels}. The ranged labels
allow the subjects to change both their grade and compartment within some range:
the grade in the numeric one and the compartment in the set one. The syntax of
these labels is 
\KBERcode{biba/effectivegrade:effectivecompartments(lograde:locompartments-higrade:hicompartments)}.
If John's subject were assigned the \KBERcode{biba/10(10-10)} label, there would be
no way for him to access the document labeled \KBERcode{biba/2}. But if the subject's
label were \KBERcode{biba/10(2-10)}, he would be able to downgrade it to
the necessary grade of 2 to read the document. For the case of the considered example,
this situation is avoided. Instead, we use the notion of a \emph{trusted entity}.
This entity is a subject that can promote the document integrity grade,
and is trusted by the higher integrity subject. In our example, this subject's user
should be able to verify the documents submitted by the subjects with the lower integrity, 
and if possible and needed, to promote the document integrity grade.
Jane and Paul are assigned these privilege and responsibility, and are assigned the
label of \KBERcode{biba/5(2-10)}. Thus, Jane can downgrade her integrity grade to
2 to read Mary's report, decide that it can be trusted and is worth for John to see, 
and then copy it to John while promoting its integrity 
grade to 10. This is achievable by the following sequence of actions:

\begin{itemize}
        \item Mary creates a \KBERcode{Report1} file in the 
        \KBERcode{UAccountingReports} folder.
        \item She moves the file to the \KBERcode{Temp} folder.
        \item Jane's current integrity label is \KBERcode{biba/5(2-10).} 
        She downgrades it to \KBERcode{biba/2} using a command like \KBERcode{setpmac biba/2 sh}, 
        and copies the \KBERcode{Report1} file
        to her home folder. The ownership of the document's copy belongs to Jane.
        \item Now Jane can read the report and check it for mistakes or forged data.
        \item After she decides that the document can be forwarded to John, she changes 
        its integrity label to \KBERcode{biba/10} by using the \KBERcode{setfmac biba/10 Report1} 
        command. She then promotes her own clearance to \KBERcode{biba/10} as well, and moves 
        the document to the \KBERcode{SummarySalesReports} folder. That folder has the integrity 
        label of \KBERcode{biba/10} and is accessible to John.
        \item Now John can read the document.
\end{itemize}

The described policy protects the integrity of the data, and makes sure
that the top management decisions are not affected by any unverified data.
We proceed by adding the confidentiality to the organization's information
security goals.

\section{Adding Confidentiality Levels}

As the company we consider is a technology one, it is rather possible that
it generates the intellectual property (IP). The Engineering department
(in our case, Paul and Robert) generates the company's IP. John should have
access to it, while Jane and her team should not. However, John may have
access to the more confidential information than Paul. For example,
he may be working with the company's partners on the subject of a common
patent or elaborating a shared strategic technology plan. Also, not only
should the confidential documents be kept away from the persons with the
lower clearance levels, but the persons with the higher clearance
levels should be prevented from disclosing the confidential information
they know, either knowingly or inadvertently. The policy to govern 
this kind of information flow may look like the following:
\begin{itemize}
        \item A confidential object with the higher security classification 
        cannot be read by a subject with the lower clearance.
        \item A subject with the higher clearance cannot write to an object
        with the lower security classification.
\end{itemize}

This policy matches the one provided by the Bell-LaPadula model with 
the liberal $\star$-property \cite{DBLP:journals/jcs/BellL96} (or the closely related BLP 
model \cite{DBLP:journals/computer/Sandhu93}). The Bell-LaPadula model assigns a clearance
label per every confidentiality grade in the organization. It is very
similar in its structure to the Biba model. For the considered case, there 
should be three clearance labels: Secret, Confidential, and
Public, with the latter one assigned to Jane and her team. The Bell-LaPadula
model is supported by the FreeBSD \KBERcode{mac\_mls} module (further referred to as MLS).

Also, it is still required to preserve the integrity, and do not loose
the work done in the previous section. FreeBSD allows this by providing
the capability of running multiple MAC policy modules simultaneously.
Thus, it is possible for the Biba and the Bell-LaPadula policies to
be in effect at the same time, and to preserve both integrity and
confidentiality.

For the FreeBSD implementation, we map the Secret clearance to the grade of
100, Confidential to 50, and Public to \KBERcode{low}. The updated file
labels are shown in Table \ref{tab:FileLabelsForMLSModel}, and the
user labels are shown in Table \ref{tab:UserLabelsForMLSModel}.

Compare the user labels assignment for the Biba and the Bell-LaPadula
policies. There are two new user logins for the Bell-LaPadula
model: \KBERcode{John.Sales} and \KBERcode{John.Engineering}.
As per the confidentiality configuration, while being in the CEO
position, John cannot directly publish the information for the Sales and
the Engineering, for his clearance level is the highest one.
By logging in as the lower clearance subject, John will be unable to access
his documents, and this will prevent him from inadvertently (or any malicious 
software on his computer from knowingly) disclosing secret and confidential
information to the lower clearance subjects.

\section{Splitting the Competence Areas}

Jane has the aggregate data of the sales activity of the company. While
the data of every individual salesperson cannot give the whole picture,
the aggregate reveals the current state of the company sales. Thus, it
might be desirable to classify the aggregate information as Confidential.
In the considered example, simply adding an MLS grade
would not help, because the MLS grades are totally ordered, and
the new labels will conflict with the Engineering department: either the
engineers will be able to read the Sales' documents, or vice versa (this
depends on whose grade is higher). In other words, the Sales
and the Engineering should have their own hierarchies with a single
point of intersection: John should have access to the documents
of both departments. This corresponds to the set-based domination
relation: for two sets $S_1$ and $S_2$, $S_1$ dominates $S_2$ (further
written as $S_1\succ S_2$) iff $S_2\subset S_1$. Obviously, incomparable
labels are supported by this definition: $\{1,2,3\}\subset\{1,2,3,5\}$,
but $\{5,9\}$ and $\{5,6,7\}$ are incomparable.
In the considered example, three labels are sufficient: 
$S_{\mbox{\scriptsize{Sales}}}=\{1\},\;S_{\mbox{\scriptsize{Eng}}}=\{2\},$ 
and $S=\{1,2\}$, where label $S$ belongs to John, so that he can access documents
belonging to any of the departments: $S\succ S_{\mbox{\scriptsize{Sales}}}$ and 
$S\succ S_{\mbox{\scriptsize{Eng}}}.$

FreeBSD supports the set-based domination relation by the notion of
\emph{compartments}. Compartments are the sets of integers, which
can be added to the Biba or the MLS labels. In the example, the labels
are directly mapped to the compartments. The MLS parts of the labels
may be rearranged, as they are not used for distinguishing
the Engineering and the Sales any longer. Instead, the MLS grades
set the confidentiality grades within individual compartments. The
resulting folder and user labels are shown in Tables
\ref{tab:FileLabelsForCompartments} and \ref{tab:UserLabelsForCompartments},
respectively. The \KBERcode{John.Sales} and \KBERcode{John.Engineering}
logins are removed, while a new one is added: \KBERcode{John.Lower.}
This new login has the same MLS grade as Jane's and Paul's ones, and is used
for creating the documents with their confidentiality classification.

With these settings, John is the only person who accesses the secret 
information. All other users have access to the confidential information,
and none of them deals with creating publicly accessible documents. John
has access to both compartments. While it is possible for him to disclose
the information of one compartment to another one, this will not
violate the confidentiality grades, and John is considered a trusted entity
for managing the inter-department information flows.

\section{The Chinese Wall Policy}

The Chinese Wall policy first formalized by Brewer and Nash \cite{DBLP:conf/sp/BrewerN89}
is oriented toward the commercial sector. This policy provides for prevention of
information flows, which cause conflicts of interest for individual consultants
working for the same consulting company. As soon as a consultant has obtained
access to the confidential information of a client company, he or she must 
have their access rights to all other companies from this industry sector 
revoked. However, the consultant must still have access to the public 
information of all client companies. The consultant starts with the access
rights to all information of all client companies. As soon as the consultant
accesses the information of a company $C$ from an industry $I$, he is then
denied access to the confidential information of all other companies from the
industry $I$. However, the consultant's access to any information of any
company from other industries is allowed. Thus, the conflict of interest does not take 
place. The deficiencies and important enhancements of the original Chinese Wall
policy model are described by Sandhu in \cite{DBLP:journals/compsec/Sandhu92}. In the same paper,
the author demonstrates that the enhanced Chinese Wall policy can be 
represented in the Bell-LaPadula model, and is therefore a lattice-based
policy. While discussing a possible implementation of this policy on FreeBSD,
we will follow the definitions given in \cite{DBLP:journals/computer/Sandhu93}.

\begin{figure}[ht]
\centering
\psfrag{EaaR8WYHeq74GME}[cc][cc]{$\mbox{SYSHIGH}$}
\psfrag{sbGvYCyCBhoPN18Wb}[cc][cc]{$\left[1,1\right]$}
\psfrag{hWwKpYpSapezvtPBE}[cc][cc]{$\left[1,2\right]$}
\psfrag{t0CKNgt5UjzblEjut}[cc][cc]{$\left[2,1\right]$}
\psfrag{qOtXiJ7lVlTyJYOJL}[cc][cc]{$\left[2,2\right]$}
\psfrag{yCGsl3nqX1m5eDWpR5Nc}[cc][cc]{$\left[1,\bot\right]$}
\psfrag{LRUk54AWWycck_G4TqiT}[cc][cc]{$\left[\bot,1\right]$}
\psfrag{dkMy9m9CCgwgL6rIhvgC}[cc][cc]{$\left[2,\bot\right]$}
\psfrag{qDVkRyoragSrXim2wkvP}[cc][cc]{$\left[\bot,2\right]$}
\psfrag{Mrl07bWju4spovG3EkBxFnU}[cc][cc]{$\left[\bot,\bot\right]$}
\psfrag{Sxn7SnSbWKbC2on4p7LawOrV}[cc][cc]{$\mbox{mls/high:1+2+3+4}$}
\psfrag{kaefOglSUFf385_XR7}[cc][cc]{$\mbox{mls/20:1+3}$}
\psfrag{xRBNE8km2AGDuKPRkq}[cc][cc]{$\mbox{mls/20:1+2}$}
\psfrag{lfxWs6Xrbl9dA9qIaL}[cc][cc]{$\mbox{mls/20:3+4}$}
\psfrag{M5qxp30lvURZYljl7S}[cc][cc]{$\mbox{mls/20:2+4}$}
\psfrag{ssKJ_ov587y1ubbx}[cc][cc]{$\mbox{mls/10:1}$}
\psfrag{cHV1HP6OjMKr9EY0}[cc][cc]{$\mbox{mls/10:3}$}
\psfrag{BlbtzOOesbDJSpkO}[cc][cc]{$\mbox{mls/10:4}$}
\psfrag{ISH9FmBxfmyfeFRQ}[cc][cc]{$\mbox{mls/10:2}$}
\psfrag{Td9nvyi0HrF1a0g}[cc][cc]{$\mbox{mls/low}$}
\includegraphics[height=3in,width=6in]{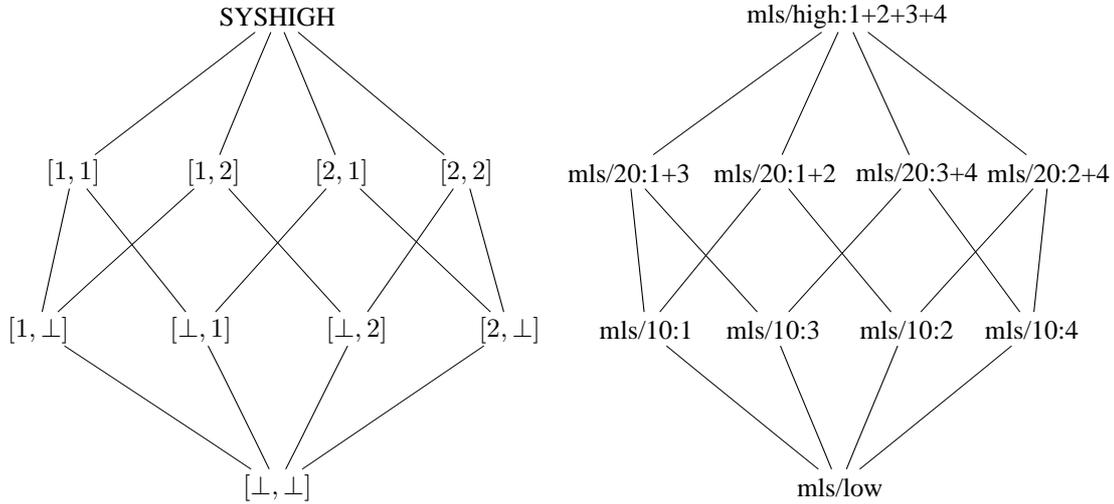}
\caption{Example Chinese Wall Lattice and Its Mapping to FreeBSD Compartments}
\label{KBER:fig_CW}
\end{figure}

Let the classes of conflict of interest (i.e. industries) be represented 
as a set $I=\left\{I_1, I_2, \ldots, I_N\right\},$ and the number of 
companies in a single industry assumed equal and be denoted by $C$. 
A clearance label will be then represented by an $N$-element vector 
$l=\{l[1], l[2], \ldots, l[N]\}, \forall{}k\in \overline{1,N}: l[k]\in I_k\cup \{\bot\}$, 
where $\bot$ in position $k$ denotes the public information of any client company 
belonging to the industry $k$ ($\bot$ is to be read as \KBERcode{null}). For example, a label for an object containing
confidential information from the company 3 in the industry 2 and the company 2 from 
the industry 4 is represented as $\left[\bot,3,\bot,2,\ldots,\bot\right]$.
There is a dominance relation for these labels defined as follows:
$l^1\geq{}l^2$ if
$\forall{}k\in\overline{1,N}:\left(l^1[k]=l^2[k]\right) \vee \left(l^1[k]\neq\bot \wedge l^2[k]=\bot\right).$
Thus, $\left[1,1,2\right]>\left[1,\bot,2\right]$ and $\left[1,1,\bot\right]>\left[1,\bot,\bot\right]$,
but $\left[1,1,2\right]$ and $\left[1,2,2\right]$ are incomparable. 
The label of all nulls $\left[\bot,\ldots,\bot\right]$ is dominated by all other labels.
The highest label is denoted by $\mbox{SYSHIGH}$. The sample lattice for the case of 2 industries and 2 companies 
per industry is shown in Fig.~\ref{KBER:fig_CW}.

It is important to understand how the lattice structure is completed 
by defining the class-combining join operator $\oplus$. Two labels 
$l^1$ and $l^2$ are called \emph{compatible} if 
$\forall k \in\overline{1,N}: l^1[k]=l^2[k]\;\vee\;l^1[k]=\bot\;\vee\;l^2[k]=\bot.$
This also means that all comparable labels are compatible, but there are
incomparable labels which are compatible, like
$\left[1,\bot,2\right]$ and $\left[\bot,3,2\right].$ Incompatible labels cannot 
be combined, while for compatible labels the following rule exists:
$l^1\oplus l^2=\left\{\forall{}k\in \overline{1,N}: \mbox{if}\;\; l^1[k]\neq\bot \;\;\mbox{then}\;\; l^1[k] \;\;\mbox{else}\;\; l^2[k]\right\}.$
For example, $\left[\bot,3,1\right]\oplus\left[2,\bot,1\right]=\left[2,3,1\right].$
We will denote the labels $l^1$ and $l^2$ as being compatible by $\hat{C}(l^1, l^2)$.
Understanding this definition will be required during feasibility analysis
for the FreeBSD implementation of the policy.

As a user progresses upwards the domination relation over the lattice,
his rights become more and more restrictive. The user leaves a trail
of subjects during this progress. For example, for user Mary, the access
to public information is granted to the subject \KBERcode{``Mary''}. 
As soon as Mary starts working on a project with a bank A, a new 
subject \KBERcode{``Mary.Banks.A''} is created for her.
When she starts working on a project with an oil company B, another subject
is created for her: \KBERcode{``Mary.Banks.A.Oil.B''}. While in a session
with \KBERcode{``Mary.Banks.A.Oil.B''} subject, Mary can read the public 
documents from the banking and oil industries, but cannot write to them.
To write to the public documents for the oil industry,
she has to close her current session and initiate a new one with the subject
\KBERcode{``Mary''}.

The assignment of the user names may be assumed to be performed
manually (for example, by a system administrator). The FreeBSD implementation 
of the Chinese Wall policy is based on this fact. All possible
login classes should be created in advance, so that adding a new user
will not lead to rebuilding the login classes. The set of login classes
required is $(C+1)^N,$ and one of those classes is a public one. The number 
of compartments required is $C\times N$. The FreeBSD MAC implementation 
imposes a limit on the number of compartments: it should be lower than 256.
This limitation should be taken into account when checking a specific organization
Chinese Wall policy for feasibility. To prevent the user from applying the 
dominance relation to the classes dominated according to the compartment 
set inclusion relation, MLS labels are added. The number of MLS grades
should be $N+1$ (taking into account the public one, and with
the exception of the $\mbox{SYSHIGH}$ class).

The rule for labeling the login classes
on FreeBSD closely follows the $\oplus$ operator considered earlier in this section.
The labeling algorithm is shown in Fig.~\ref{KBER:ChWall}. As a result
of applying this algorithm to the original lattice, a set of triplets 
is constructed. A triplet contains an MLS grade, a FreeBSD compartments set, and
a corresponding original Chinese Wall label. The algorithm iterates over the
lattice by examining the set of labels with a fixed non-$\bot$ set of positions 
on every step. The previously constructed triplets are examined for containing
the original labels dominated by the current one. The corresponding FreeBSD
compartment sets are combined and added to the currently constructed triplet. 

\begin{figure}[ht]
\begin{algorithmic}[1]
\State $L^1\gets\left\{l \;:\; \exists i: (l[i]\neq\bot) \;\wedge\; (\forall j\neq i : l[j]=\bot)\right\},\;M^1 \gets \emptyset,\; i\gets 1$
\ForAll {$l \in L^1$}
        \State $M^1 \gets M^1 \cup \langle 10, i, l\rangle$
  \State $i\gets i+1$
\EndFor
\For {$i \gets 2, N$}
        \State $g \gets \{\mbox{\bfseries if} \;\; i\neq N \;\;\mbox{\bfseries then}\;\; i\times 10 \;\;\mbox{\bfseries else}\;\;\mbox{high}\}$
        \State $X^i \gets $ set of all $i$-subsets of $\{1,\ldots, N\}, \; L^i \gets \emptyset,\; M^i \gets \emptyset$
        \ForAll {$x \in X^i$}
                $L^i \gets L^i \cup \left\{l \;:\; (\forall j \in x : l[j]\neq\bot) \wedge (\forall j \notin x : l[j]=\bot), \; j=\overline{1,N}\right\}$
                \Statex Now $L^i$ contains all lattice node labels with $i$ positions being not $\bot$.
        \EndFor
        \ForAll {$l \in L^i$}
                \State $M'\gets\{\forall m'\in M^{i-1}: m'=\langle g',f',l'\rangle \wedge \hat{C}(l,l')\}$
                \State $M^i\gets M^i\cup\{\langle g, \bigcup_{m'\in M'}f', l\rangle\}$
        \EndFor
\EndFor
\State $M\gets{\bigcup}_{i} M^i$
\end{algorithmic}
\caption{FreeBSD Label Generation for Chinese Wall Policy}
\label{KBER:ChWall}
\end{figure}

The system 
administrator has to create a set of login classes for the combined 
Bell-LaPadula and compartment labels.
When a user starts working with a company from a new industry, the administrator
provides the user name corresponding to the changed Chinese Wall lattice label.

\section{Conclusion}

The set of the policies implemented by the FreeBSD/TrustedBSD project allows
to implement not only the Bell-LaPadula and Biba models. By reusing the 
result of the Chinese Wall model being a lattice-based one, it can be
implemented in the FreeBSD MAC framework as well. Also, the level of MAC 
support allows implementing trusted entities, so that the information
flows are possible both ways, but in a controllable manner, so that the
policies are not violated. The combination of a number of MAC models
provides a very flexible platform for satisfying the needs of
various organizations. The authors hope that the proper illustration of
FreeBSD MAC support capabilities will promote the usage of mandatory
access control in the commercial sector.

\section*{Acknowledgment}

The authors are grateful to Alexey Fomin whose comments and suggestions
have significantly improved the paper.

\bibliography{IEEEabrv,KBERbibfile}

\end{document}